\documentclass[11pt]{article} 
\usepackage{moriond,epsfig}

\def\Journal#1#2#3#4{{#1} {\bf #2}, #3 (#4)} 
 
\def\NIMA{{\em Nucl. Instrum. Methods} A} 
 
\def\PLB{{\em Phys. Lett.}  B} 
\def\PRL{\em Phys. Rev. Lett.} 
\def\PRD{{\em Phys. Rev.} D}

\def\be{\begin{equation}} 
\def\ee{\end{equation}} 
\def\bea{\begin{eqnarray}} 
\def\eea{\end{eqnarray}} 
 
\begin{document} 
\vspace*{4cm} 
\title{RESULTS OF THE PIERRE AUGER OBSERVATORY ON ASTROPARTICLE PHYSICS} 
 
\author{ S. ANDRINGA } 
 
\address{LIP, Av. Elias Garcia, 14, 1,\\ 
1000-149 Lisboa, Portugal\\ 
{\bf for the PIERRE AUGER COLLABORATION}} 
 
\maketitle\abstracts{ 
The Pierre Auger Observatory has already collected more ultra high energy cosmic ray data than all the previous experiments. With an hybrid detection technique, it can provide coherent results on the flux, energy spectrum and arrival directions of the highest energy cosmic rays, and characterize the extensive air showers in order to probe the primary particle characteristics and its interactions. These results will be presented from the point of view of particle physics.} 
 
\section{Introduction} 
 
High energy cosmic rays may give important information on different areas of Physics. The sources where they are produced, the corresponding acceleration mechanisms, and even their precise composition in terms of primary nuclei remain unknown and are important questions from the astrophysical point of view. When they arrive at Earth, they collide with the atmosphere nuclei at center-of-mass energies which are orders of magnitude above the ones available in man-made particle accelerators, allowing the probe of particle physics at a new energy scale. In between, they propagate through the interstellar and intergalactic media, subjected to magnetic fields and to interactions with cosmic matter and radiation, having the possibility to give indirect indication about phenomena which arise only at large distances and energies. 
 
The Earth atmosphere acts as an efficient calorimeter for the high 
energy cosmic rays. The initial particle interacts with the 
atmospheric nuclei creating large numbers of lower energy particles 
in a showering process. Nuclei will produce charged pions which can 
interact or decay, producing both the atmospheric neutrino flux and 
high energy muons that can be efficiently sampled at ground. Neutral 
pions will decay immediately into gammas feeding an electromagnetic 
shower which will carry most of the initial energy. The 
electromagnetic shower will produce isotropic fluorescence light by 
exciting the nitrogen molecules in the air which allows the imaging 
of the showers passing in front of a telescope regardless of their 
direction. The intensity of this light is directly proportional to 
the deposited energy. 
 
\section{The Pierre Auger Observatory} 
 
\begin{figure} 
\includegraphics[width=0.5\textwidth]{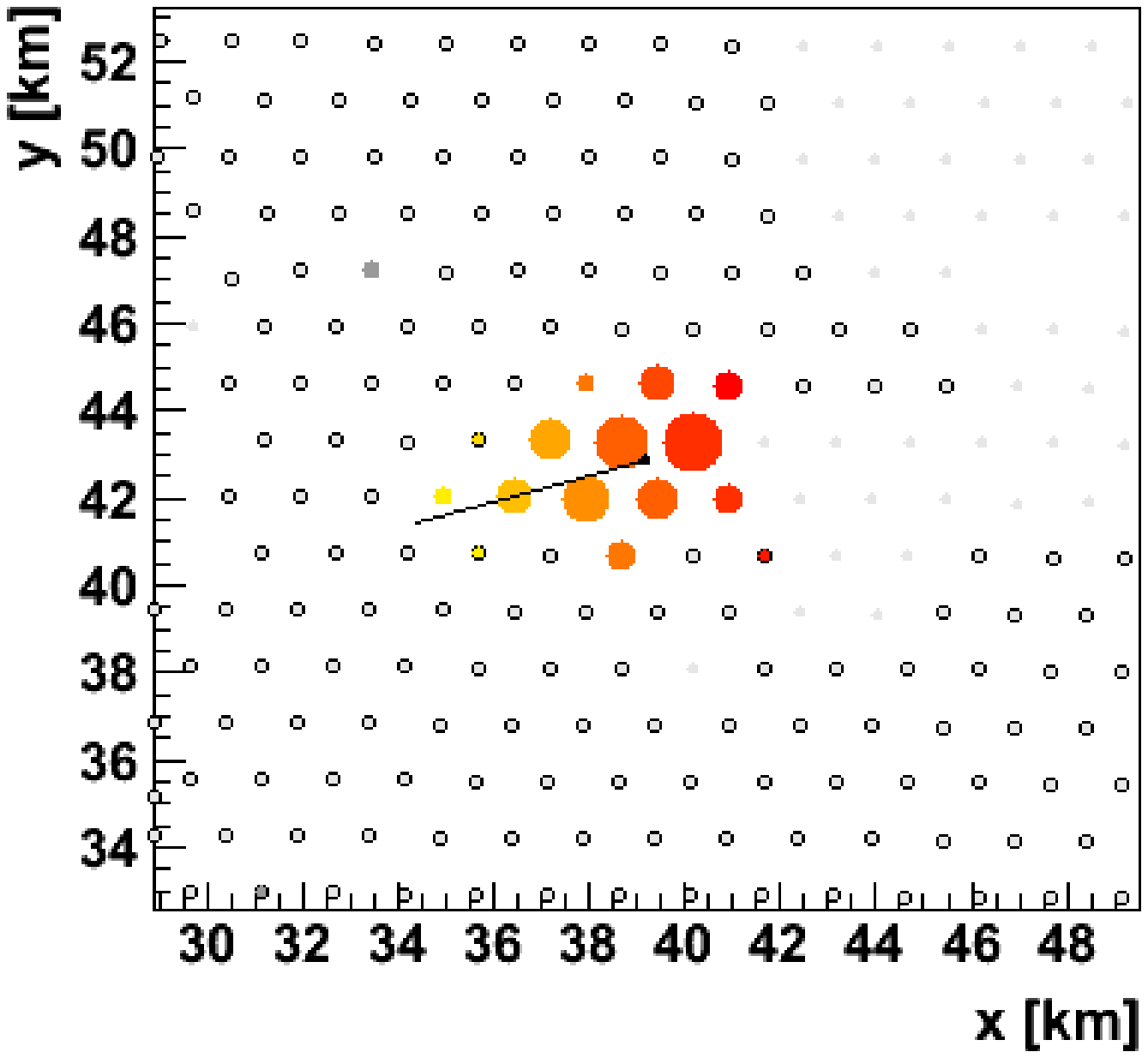} 
\includegraphics[width=0.5\textwidth]{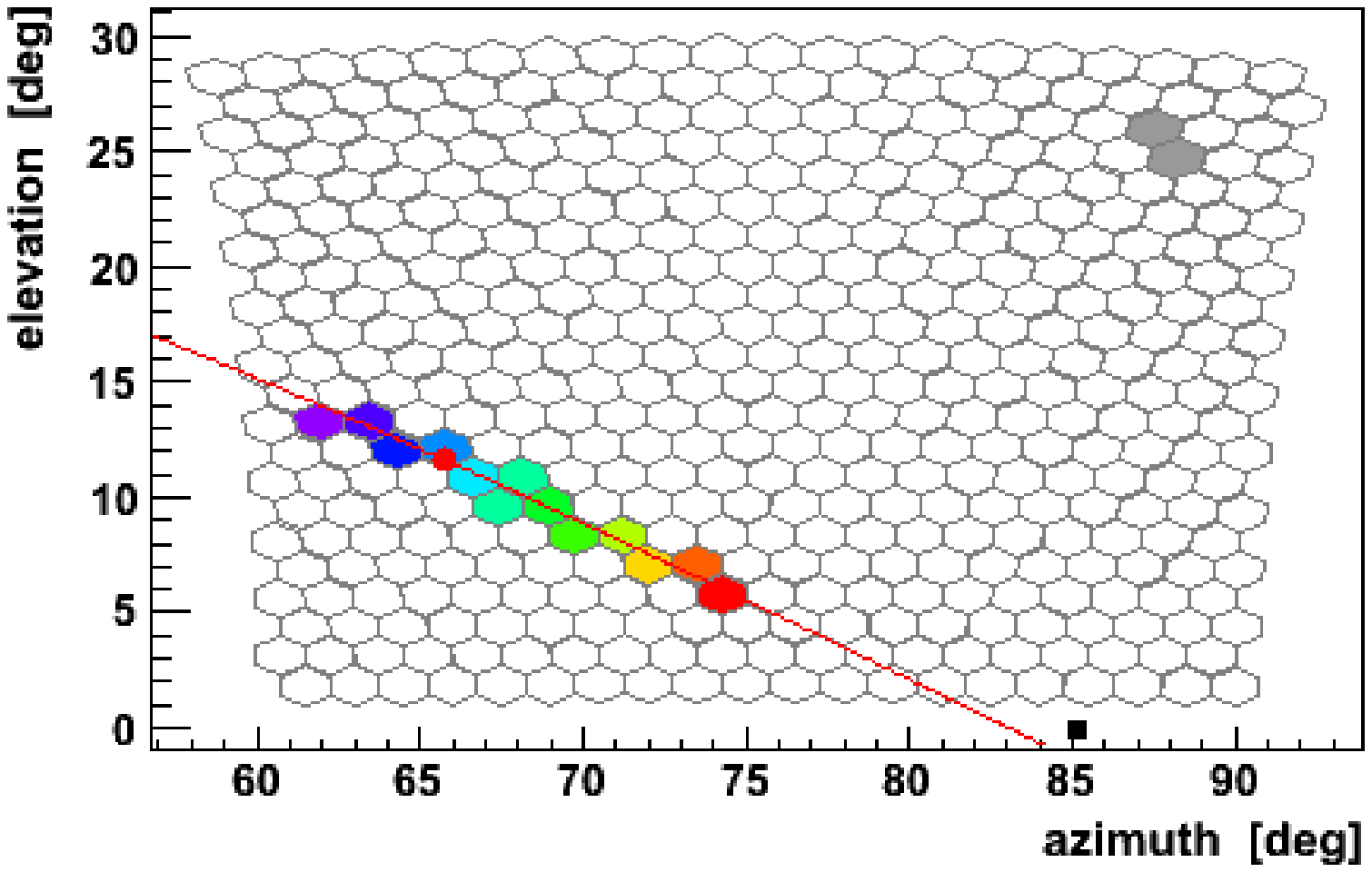} 
\includegraphics[width=0.5\textwidth]{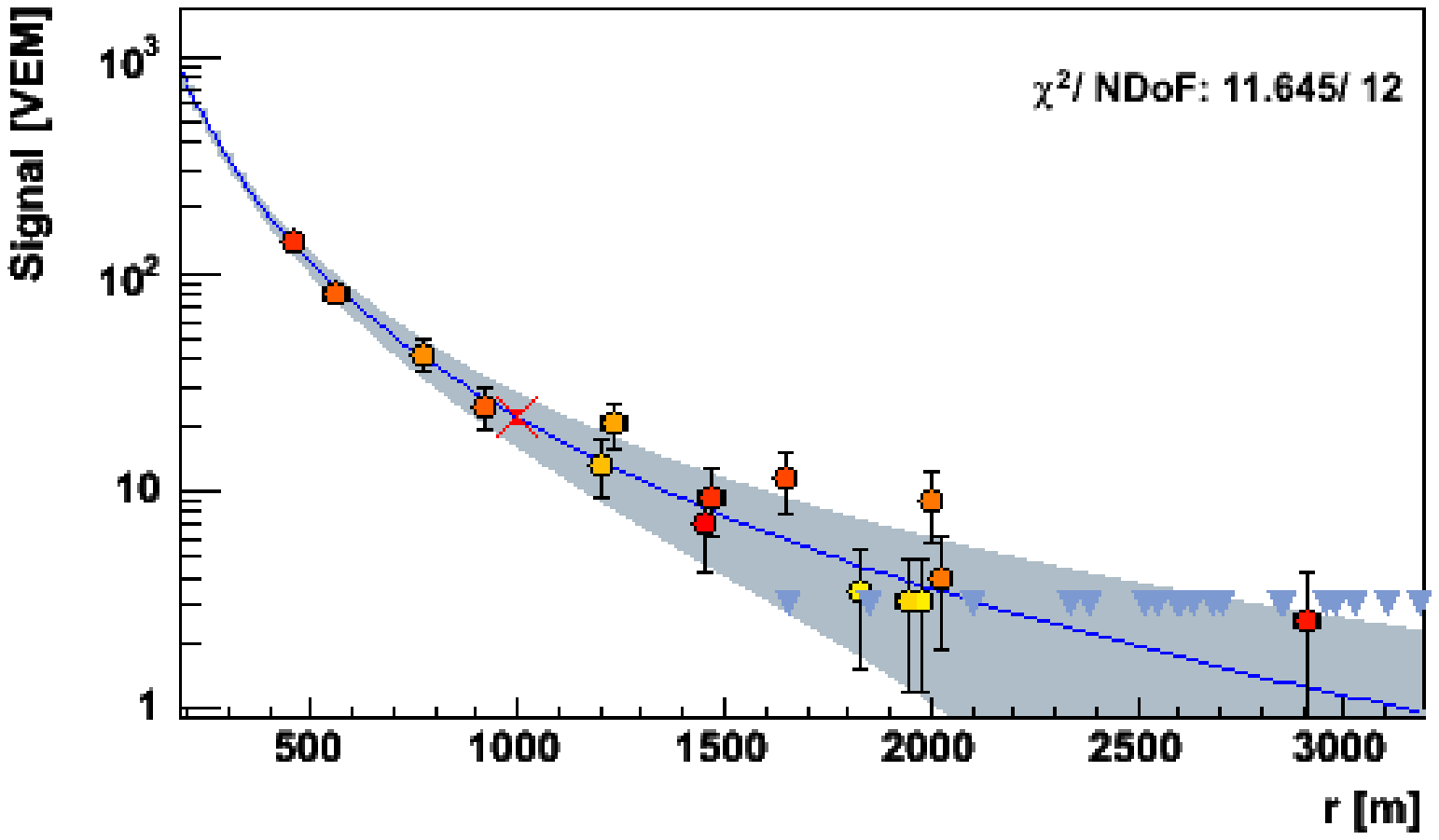} 
\includegraphics[width=0.5\textwidth]{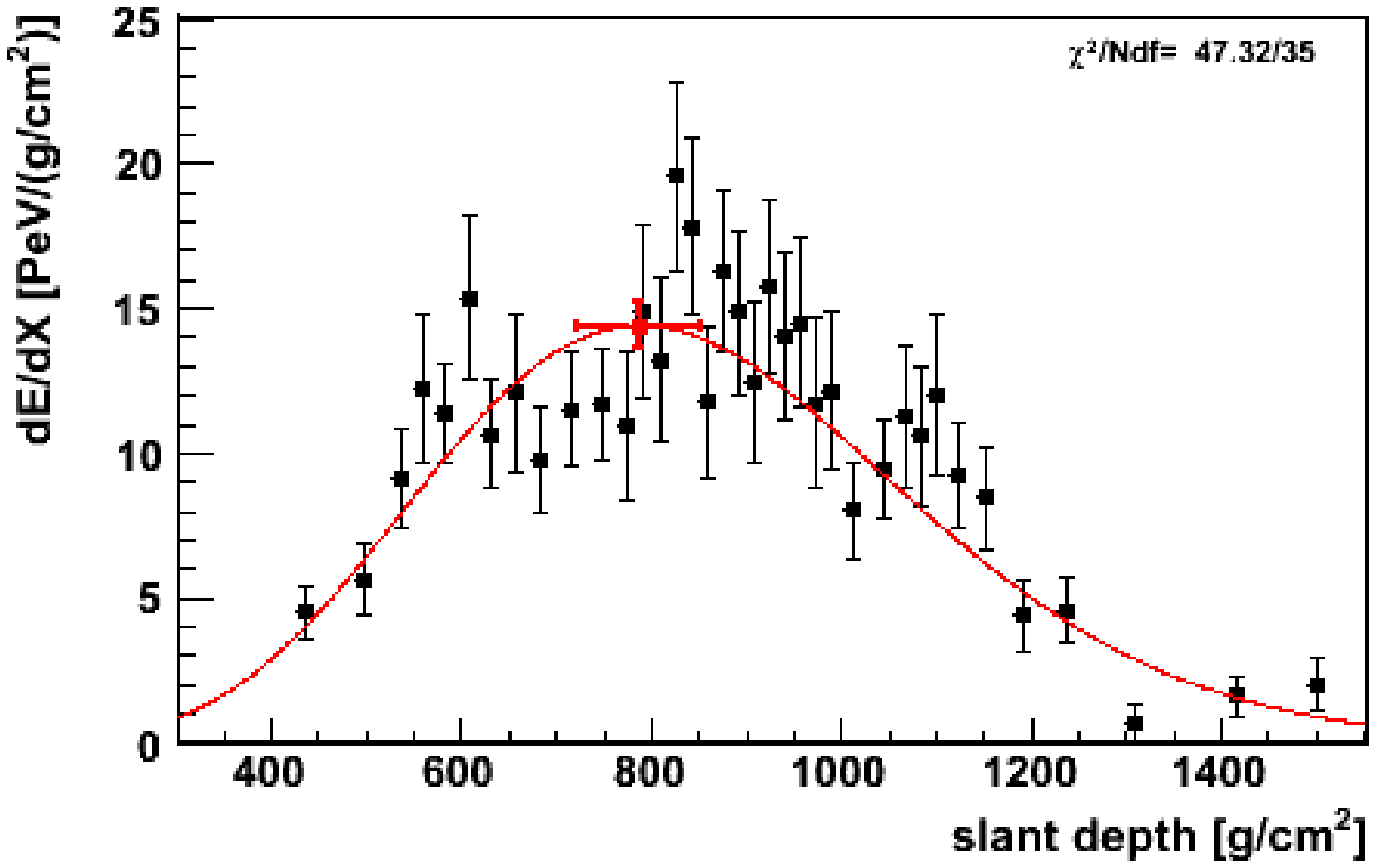} 
\caption{The same event seen by the Surface Detector (on the left) and the Fluorescence Detector (on the right). The colors indicate timing of arrival of particles to the SD tanks, or of light to the FD camera. The two lower plots show the SD lateral profile, measured as number of particles as a function of distance to the core, and the FD longitudinal profile of energy deposition as a function of atmospheric depth.\label{4eyes} 
} 
\end{figure}

The Pierre Auger Observatory~\cite{observatory} is the largest 
operating project to measure cosmic rays of energies above 
10$^{17.5}$ eV. Data has been collected from the start of 
construction, in 2004, and amounts now to the equivalent data that 
will be collected in two years of the full Observatory. The 
Observatory is a hybrid detector, where cosmic ray induced 
air-showers are detected both by particle sampling on the ground and 
fluorescence light imaging. The Surface Detector covers 3000~km$^2$, 
by means of 1600 pure water Cherenkov tanks, covering around 
10~m$^2$ each and separated by 1.5~km. Each individual surface 
detector station has its own solar panel and battery and GPS for 
time synchronization. 3 photomultiplier tubes record the passage of 
charged particles in the tank with 25~ns sampling, and the results 
are transmitted by radio to the closest Fluorescence Detector site. 
There are four sites with 6 telescopes each, observing the 
atmosphere above the array. The information is read in 100~ns time 
bins by 440 pixels, covering elevations from 2 to 32 degrees. Cosmic 
rays arriving in moonless clear nights are seen both by the Surface 
Detector (SD) and the Fluorescence Detector (FD), and even by 
several of the FD eyes in some cases. The possibility of 
simultaneous measurements of the different components of the same 
cosmic ray air shower is one of the most important characteristics 
of the Observatory. These events are used for calibration of the 
surface detector energy estimator, for correlation of indirect 
measurements of the depth of maximum shower development and for 
estimating the resolution of the reconstructed variables. An example 
event is shown in fig.\ref{4eyes}. 
 
The relative timing between SD stations allows the precise 
reconstruction of the incident cosmic ray direction. The particle 
density falls rapidly as the distance from the shower core 
increases, and differently according to different shower models. The 
density at 1~km, corrected for different attenuation as a function 
of zenith angle, was found to be a stable energy estimator. The 
signal in each tank contains both a relatively smooth 
electromagnetic contributions and more peaked signals due to the 
arrival of muons - far from the core, the muon component dominates 
and can be estimated. 
 
The direction reconstruction for the FD is based on the image and 
timing, and complemented by the timing of the highest signal SD 
station. The shower development as a function of the crossed 
atmospheric depth is measured almost directly, and can be 
parametrized empirically. The integral of this function is a direct 
calorimetric energy measurement while the depth at which the shower 
maximum is reached ($X_{max}$) is an important parameter to 
distinguish different cosmic ray primary particles. Also a Cherenkov 
cone accompanies the shower and, even if it does not point at the 
eye, it can be scattered and contaminate the fluorescence signal. 
The profile function is extracted taking into account that both 
contributions are present and originate from the same shower 
particle distributions. 
 
The calibration of the SD energy estimator with FD data allows the 
Pierre Auger Observatory to combine the very large exposure to a 
small systematic uncertainty on the measured energies. The present 
systematic uncertainty is of 22\%, dominated by uncertainties in the 
fluorescence yield, absolute FD calibration and reconstruction and 
in the atmospheric attenuation and scattering. The atmosphere is 
monitored with several redundant techniques, and its evolution in 
time is taken into account in event reconstruction. The angular 
resolution obtained for the arrival directions reconstructed with 
the SD is better than 1 degree. 
 
\section{High Energy Cosmic Ray Spectra, Fluxes and Arrival Directions}

\begin{figure} 
\begin{minipage}{0.5\textwidth} 
{ 
\includegraphics[width=0.99\textwidth]{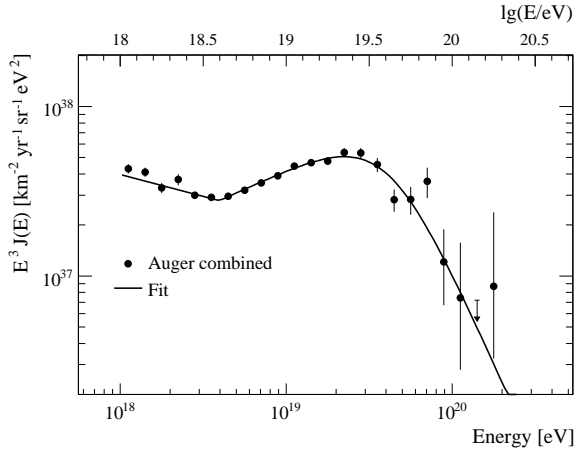} 
} 
\end{minipage} 
\hfill 
\begin{minipage}{0.5\textwidth} 
{ 
\caption{The figure shows the measured cosmic ray flux multiplied by E$^3$. The FD and SD data were checked to be compatible and are combined. \label{spectra_map} 
%The right figure shows the arrival directions of the 58 cosmic ray events with energies above 55 EeV, in galactic coordinates. The shaded area represents the AGN density obtained by smearing the Swift-BAT AGN catalog and folding it with the observatory exposure. 
} 
} 
\end{minipage} 
 
\end{figure} 
 
The cosmic ray flux measured as a function of energy is shown in 
fig.\ref{spectra_map}. The result~\cite{spectra} combines the 
information of the FD (extending the efficiency to energies below 
the log(E/eV)=18.4 SD threshold) and the SD, which dominate due to 
higher statistics. The flux is multiplied by $E^3$, to compensate 
the steep falling power law spectra and put in evidence two features 
at log(E/eV)=18.6 and log(E/eV)=19.5. The first break is known as 
the {\it ankle} and can be interpreted as marking the transition 
between galactic and extra-galactic cosmic-ray origins. While lower 
energy cosmic rays are trapped by the galactic magnetic fields, the 
high energy flux may be dominated by particles that originated in 
very distant sources. As the energy increases their trajectories 
will be less and less affected by magnetic fields, and they will 
start pointing back to the sources. Astronomy might thus become 
possible. The second break could be due to the end of accelerating 
power at source, but it has been predicted long ago that cosmic ray 
fluxes at these high energies would have to be attenuated by 
collisions with the Cosmic Microwave Background (CMB). When the 
center-of-mass energy of those collisions is above roughly 
1~GeV/c$^2$, they lead to nuclear disintegration and photo-pion 
emission from the excitation of proton resonances. Most nuclei can 
only travel a few tens of Mpc. Even pure beams of the very stable 
iron nuclei or proton primaries would be attenuated to 50\% after 
100 Mpc. The effect described for protons is know as the GZK 
cut-off~\cite{GZK}.

Most of the high energy cosmic rays detected arrive isotropically, 
their directions modulated only by the detector exposure. For the 
highest energy data, however, the arrival directions show a 
correlation with the locations of the nearest extra-galactic 
potential sources. An excess of events is seen close to Centaurus A, 
the closest Active Galaxy Nuclei (AGN): within $18^\circ$, 12 events 
are seen, while only 2.7 were expected from an isotropic 
distribution~\cite{correlnew}. 
 
The first evidence for anisotropy~\cite{correlold} came from a 
comparison of data with the V\'eron-Cetty and V\'eron (VCV) AGN 
catalog. The correlation was maximised by optimizing three 
parameters: a maximum angular distance of $3.1^\circ$, an energy 
threshold of 55 EeV, and a maximum source distance of 75 Mpc. These 
parameter values were fixed with a small number of events. With four 
times that initial exposure, 58 events are found above the energy 
threshold, and the data continue to show an excess of correlation in 
respect to an isotropic distribution, even if lower than in the 
initial data-set. The probability that the new data arises from an 
isotropic distribution is 0.006~\cite{correlnew}. Likelihood tests 
and cross-correlation analysis with other catalogs indicate that the 
directions are partially correlated with the distribution of local 
matter~\cite{correlnew}. 
 
While correlation parameters could be interpreted as an indication of 
small magnetic deflection and the GZK cut-off, the exact sources or 
the charge of the primary particles are still not determined. 
 
\section{Primary Particle Identification} 
 
The indirect indications on the primary particle types of the cosmic rays given by the spectrum and direction results are, in any case, not enough if one wants to use them for particle physics. It is their behavior at detection in the atmosphere that will show what kind of particle really arrived - which can be the one emitted by the initial source or result of further interaction during propagation. Particles such as neutrinos, photons, protons or nuclei will produce different signals which may be distinguished even when the precise high energy interactions are unknown, since the main differences arise from the total interaction cross-sections in air. 
 
\begin{figure} 
\includegraphics[width=0.5\textwidth]{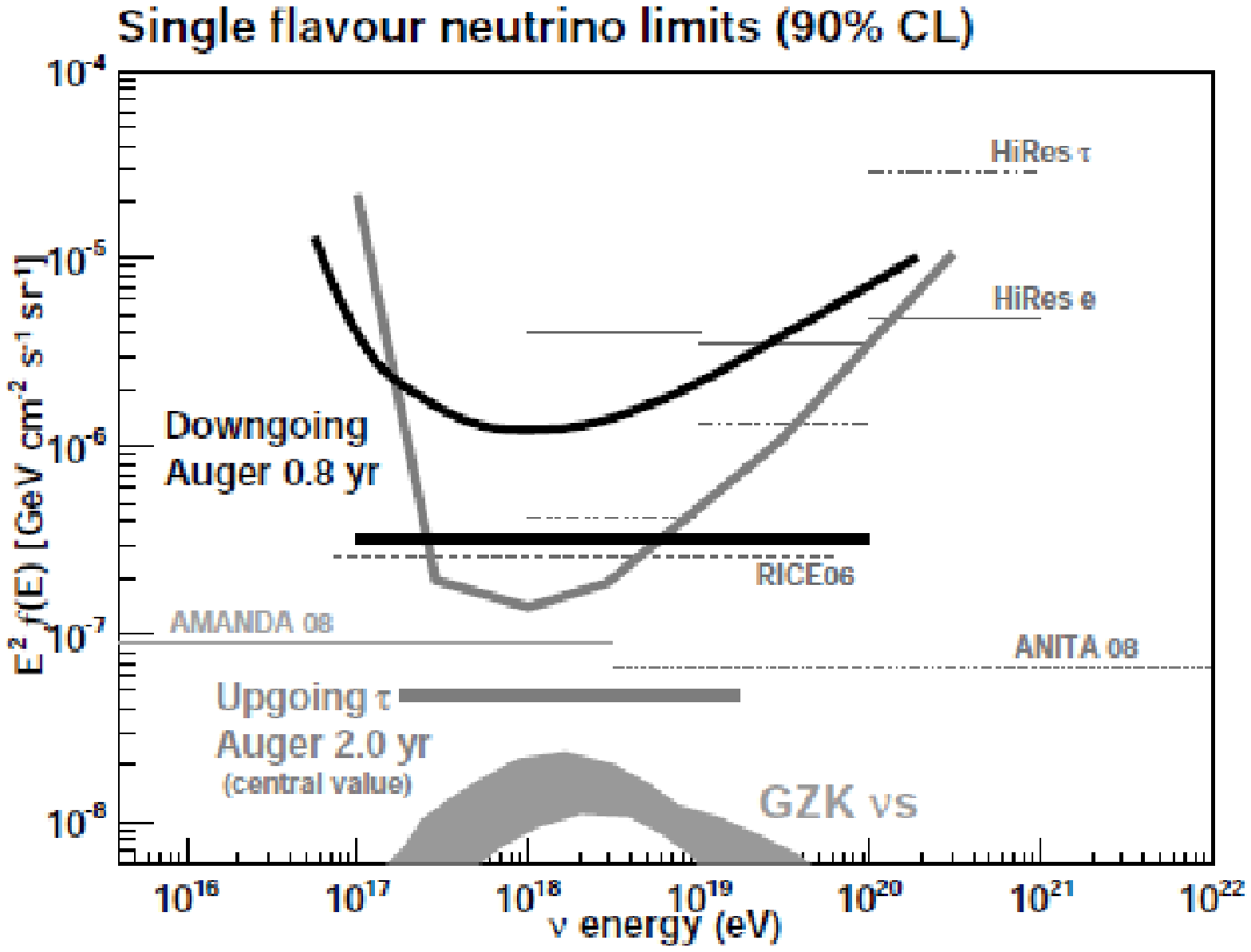} 
\includegraphics[width=0.5\textwidth]{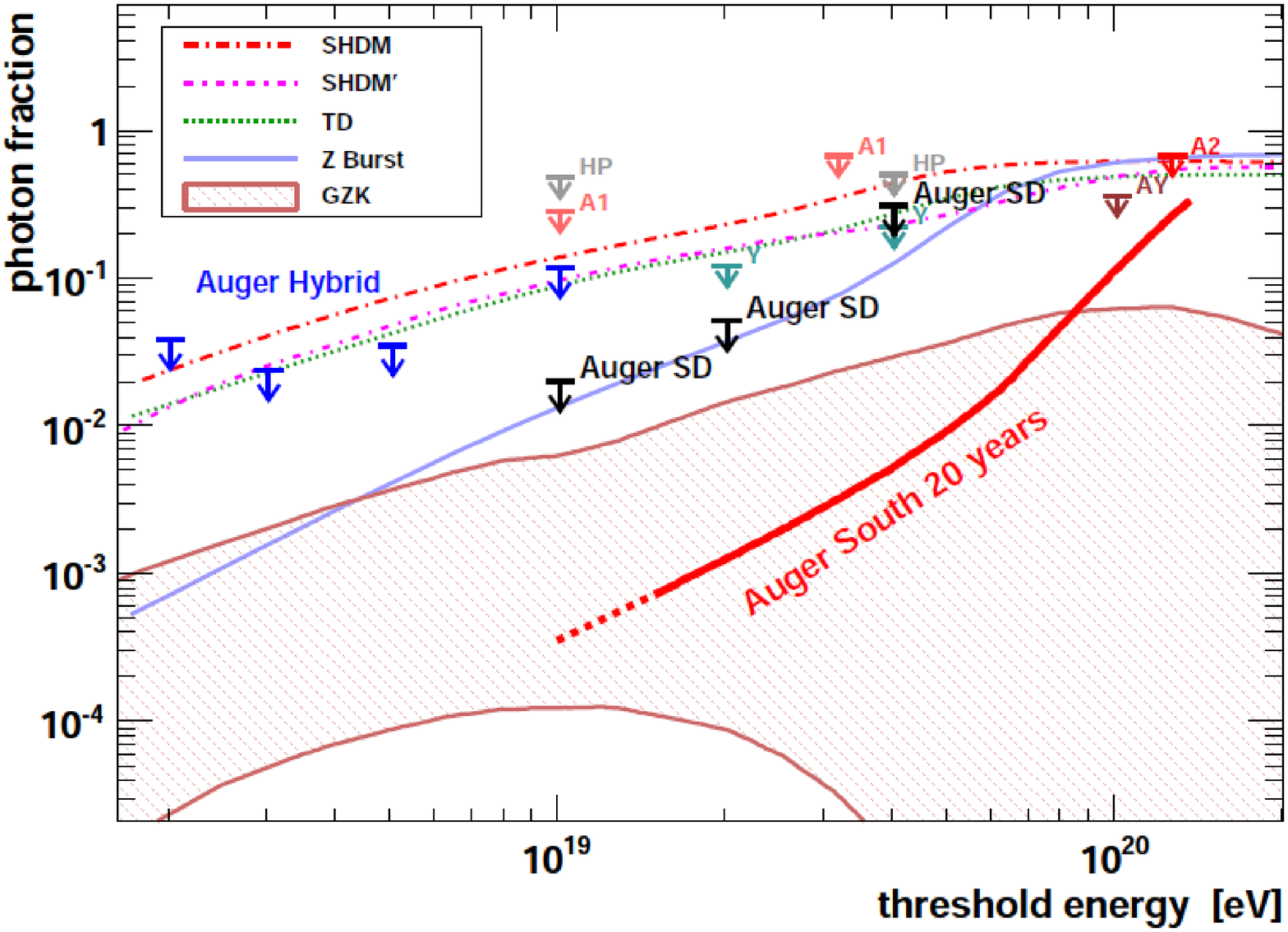} 
\caption{Limits on the absolute flux of neutrinos (left) and on the fraction of photons relative to the total cosmic ray flux (right). 
Results from the several Auger analysis are compared to those of other experiments, and to predictions from proton interactions with the CMB. 
\label{limits} 
} 
\end{figure}

Neutrinos will very rarely interact in the atmosphere. The total vertical depth of the atmosphere over Auger is less than 1000 gcm$^{-2}$, but showers coming from horizontal directions can have crossed 20 times larger depths. In that case, most of the electromagnetic component will have been absorbed, and only muons will survive. Only neutrinos will cross those depths without interacting and create young horizontal showers, with an important electromagnetic component. Moreover, tau neutrinos can be seen as Earth skimming showers initiated by energetic taus emerging from the Earth or the Andes mountains, in an almost background free signature. Unfortunately, none of these spectacular signatures was found up to now, and limits were set on the allowed absolute fluxes~\cite{neutrino}. These limits are comparable to those from dedicated high energy neutrino experiments, with a peaked sensitivity at energies around 1 EeV, as shown in fig.\ref{limits}. 
 
Less spectacular but still distinguishable signatures are expected 
for primary photons, again due to the absence of strong 
interactions. Showers initiated by photons can be described almost 
precisely from pure Quantum Electrodynamics. The first photon will 
produce an electron-positron pair, which will then emit more 
photons, repeating the process until the energy of each particle is 
too low and the cascade dies out. The maximum number of particles 
achieved depends on the the initial energy. The depth at which 
maximum occurs can have large fluctuations. At the energies to which 
the Pierre Auger Observatory is sensitive, shower maximum is reached 
close to the ground. Deep profiles can be searched for in the 
Fluorescence Detector data, while in the Surface Detector they will 
be seen as curved shower fronts with very wide time signals in the 
tanks, characteristic of the electromagnetic component. In this 
case, the search is not background free and the limit shown in 
fig.\ref{limits} is set on the allowed fraction of primary photons 
as a function of energy~\cite{photon}.

Neutrino and photon primaries had been proposed before to explain the trans-GZK cosmic ray fluxes, and the data allows us to constrain most of those models in which they would arise from heavy relic particles decay. However, even if they can not be accelerated they are expected to be produced at the sources, due to the interactions and decays of the charged particles, and could be very useful for precision astronomy and study of the sources. On the other hand, if the decrease of the high energy fluxes is caused by interaction with the CMB, then both neutrinos and photons are expected from pion decays, produced during charged cosmic ray propagation. In fig.\ref{limits}, these predictions are also shown: in ten years, those signals should be seen by the Pierre Auger Observatory, and the relation between the fluxes of different primaries could then help to solve the puzzle of the charged flux. 
 
Most of the highest energy cosmic rays must then be protons, iron or 
intermediate mass nuclei. A sensitive observable to distinguish 
between them is the depth of maximum of the electromagnetic 
shower~\cite{xmax}. This maximum depth increases with energy but 
will occur earlier for protons than for photons, both due to the 
higher cross-sections and to the simultaneous creation of many 
secondary photons already at the first interactions; in the same way 
also iron initiated showers will in general penetrate less than 
proton ones. For events seen by the FD, the precise energy and 
$X_{max}$ of each event are both obtained by a fit to the 
longitudinal profile, resulting in resolutions of 22\% and 
20~gcm$^{-2}$, respectively. 
 
\begin{figure} 
\includegraphics[width=0.5\textwidth]{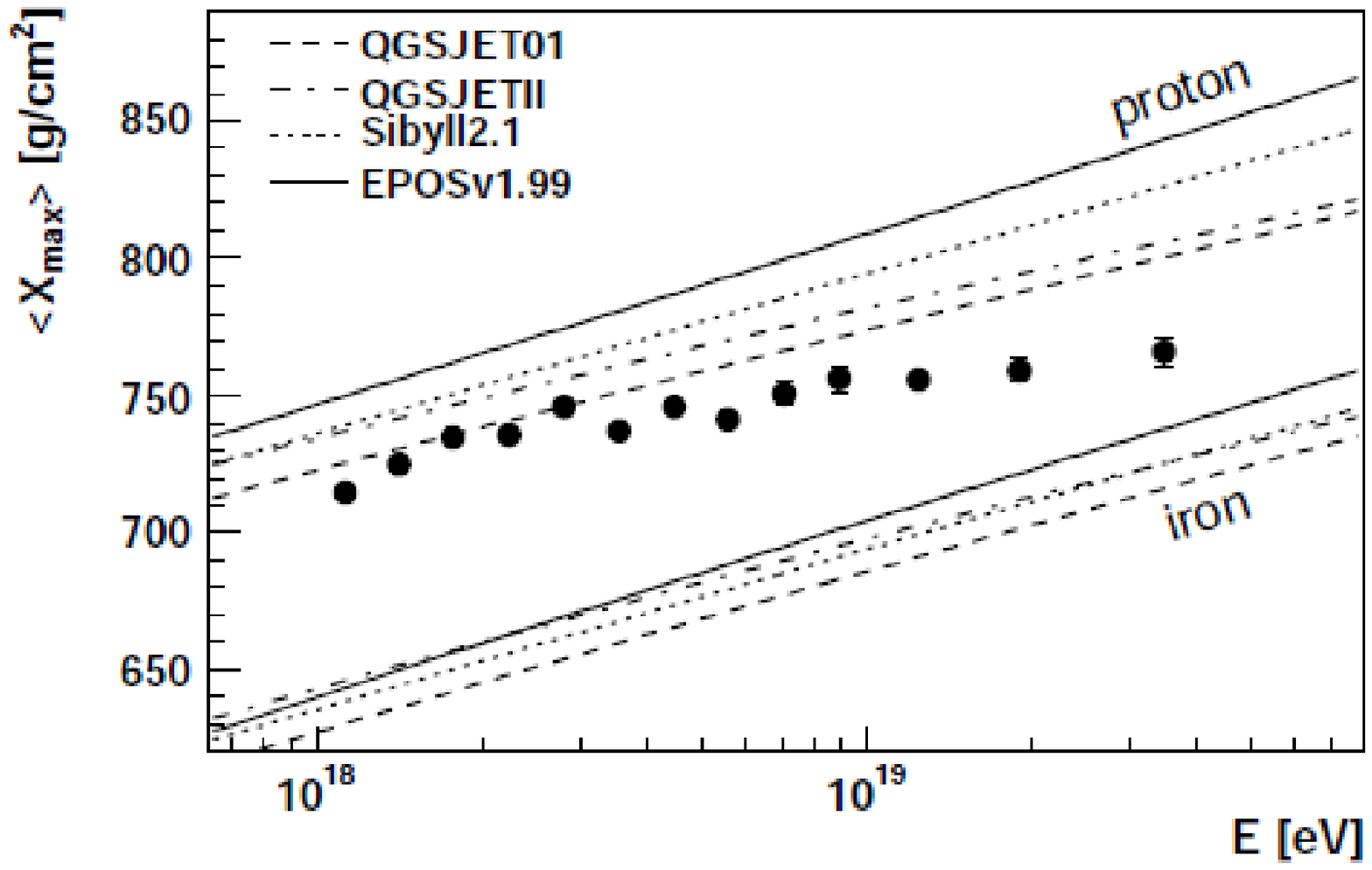} 
\includegraphics[width=0.5\textwidth]{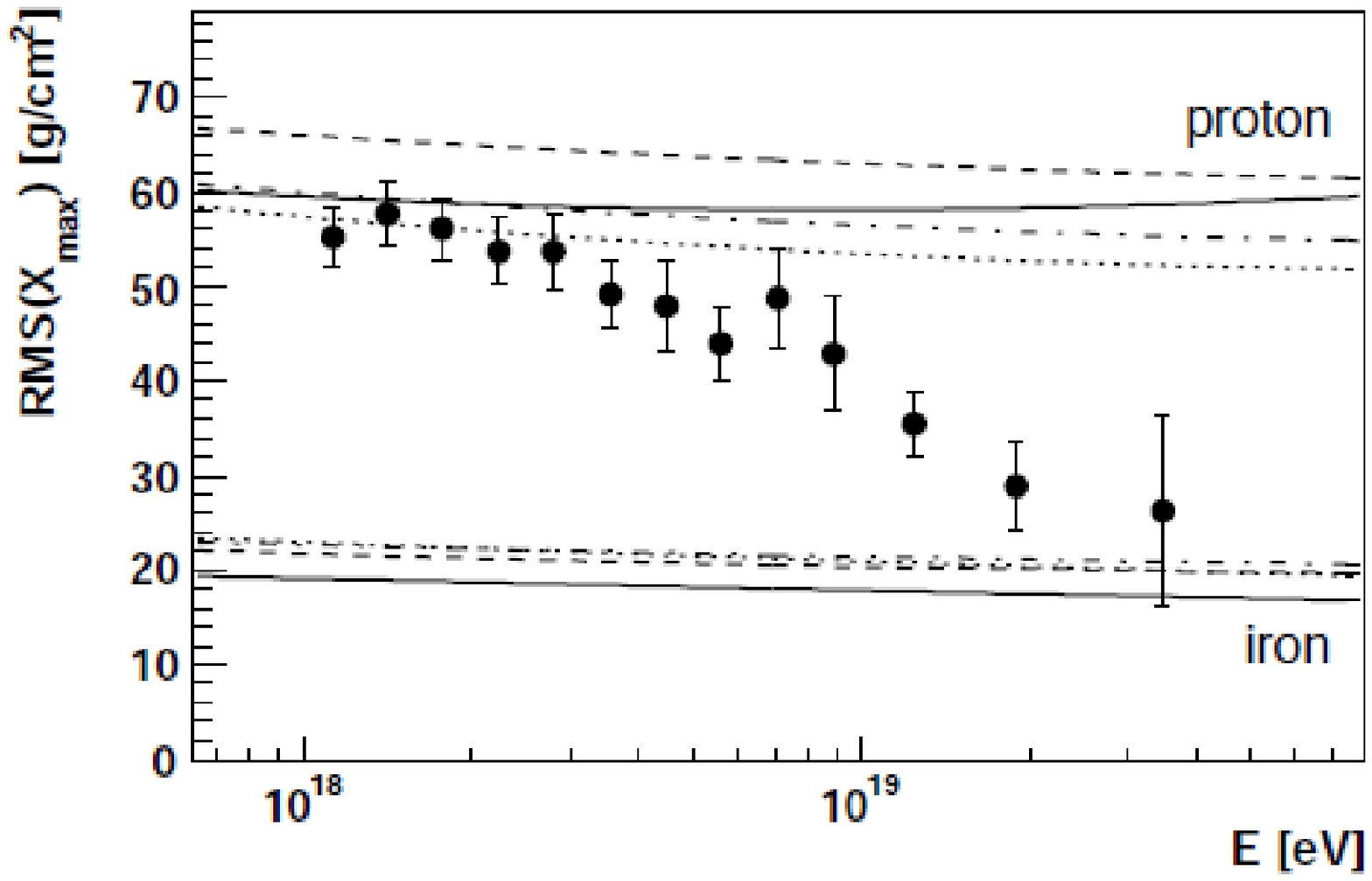} 
\caption{The mean (left) and dispersion (right) of the shower maximum distribution obtained by the FD as a function of energy. 
The predictions of different hadronic models, for pure proton and iron composition, are also shown. \label{xmax} 
} 
\end{figure}

Fig.\ref{xmax} shows the evolution of the mean $X_{max}$ with energy compared to the predictions of different hadronic models. While the low energy data is reasonably compatible with the expectation for protons, at around log(e/eV)=18.25 the data deviate from the expectation, approaching the prediction for heavy nuclei, indicating a mixed composition with average mass in between proton and iron. An extension of these results to higher energies, and in particular to the ones where we observe the sudden flux decrease and anisotropy, can at present only be done with the higher statistics Surface Detector data, in indirect ways. Preliminary analyzes of different variables show that the trend continues and the data becomes more similar to iron~\cite{xmax_sd}. 
 
This measurement alone is not enough to separate a change in composition from a change in the hadronic interactions. The interpretation can be that the proton-air interaction cross-section or multiplicity, or both, are increased leading to a faster shower development, as would happen in a heavy nuclei collision.  The full distribution of $X_{max}$ can bring more information, and that analysis is now being performed at the Pierre Auger Observatory, for the first time. As shown in fig.\ref{xmax}, not only the mean of the distribution, but also the width shows the same unexpected behavior, in an even clearer way. The last energy point available is even compatible with a pure iron composition, that is, small fluctuations in the first interaction point, but also in the subsequent shower development. Since there is only one clear break in the distribution, it is possible to think of a smooth composition transition, or a cumulative effect of the energy evolution of the cross-section 
 s of a single primary, but it is interesting to note that the energy at which the data becomes incompatible with pure proton composition is close to the transition between galactic to extra-galactic dominated fluxes. 
 
The Surface Detector allows also for an indirect count of muons in the showers, and again several different methods were tested and agree with each other. Muons are interesting to isolate since they can bring information directly from the first hadronic interactions. The data seems to have too many muons compared to the expectations for proton showers, and is even slightly above the iron expectations. Once again, however, these results can not be interpreted as an absolute proof of the cosmic ray composition, as they can be due to changes in the high energy hadronic interactions. In fact, to date, none of the available hadronic models can fully describe the data, with or without a composition change~\cite{xmax_sd}.

\section{Towards a Full Observatory for Particle Physics} 
 
Clearly, it would be desirable to fully characterize the primary beam before particle physics results can be extracted from cosmic rays, but probably composition and interaction studies will have to be done simultaneously. The fact that the lower energy data is compatible with a pure proton description might allow for a more detailed study at these center-of-mass energies - at and above the LHC - and allow for the tuning of the different model parameters in a consistent way. Data from the LHC will also constrain the energy evolution. In particular the results from dedicated forward physics LHC experiments will help in constraining the evolution for cosmic ray energies up to around 10$^{17}$ eV. The interplay between cosmic rays and accelerators results will improve the studies of the next energy frontier. 
 
The search for rare and unexpected physics has always been an 
important task in the history of cosmic rays. Historically, the main 
aims were not merely observing the primary particles arriving at 
Earth, but on seeking new, unknown particles produced in the air 
showers. Even if cosmic ray detectors provide less information than 
typical particle physics detectors, the detailed study of the shower 
evolution can give information of new physics. The luminosity is 
low, and so only high cross-section processes or new properties 
affecting consistently the full shower development, are expected to 
be observable. The detailed study of the ``normal'' shower 
characteristics and of their fluctuations must be complemented with 
a detailed study of the detector systematics, which include besides 
the detector instruments themselves also the atmosphere above the 
Observatory, monitored regularly with a large number of redundant 
techniques~\cite{atmosph}. 
 
The Pierre Auger Observatory in the Southern Hemisphere is being 
extended in order to give a more complete view of the highest energy 
cosmic rays with respect to a particle physics point of view. Lower 
energy extensions~\cite{lowenergy} will include infill arrays, where 
the spacing between the Surface Detector tanks is reduced in order 
to probe part of the showers in more detail, and to be able to 
detect lower energy cosmic rays, so has to be able to overlay with 
lower energy observatories. The new tanks will be accompanied by 
buried scintillator detectors for muon identification, so that the 
degeneracy with the electromagnetic component can be broken. 
Although the infill is done only in a smaller area, the flux 
increases enormously at lower energies, so the statistics will be 
comparable. In the same small area, the Fluorescence Detector will 
be extended to higher elevations. At lower energies the shower 
reaches its maximum higher in the atmosphere, and that maximum will 
be thus recorded. 
 
At the same time, the Observatory is preparing a new site in the 
Northern Hemisphere~\cite{north}. This will be dedicated to the 
highest end of the spectrum, covering an area roughly seven times 
larger than the existing Southern Observatory, with a slightly 
larger distance between tanks. This will allow full sky coverage 
with the same hybrid technique, for the first time. 
 
\section{Conclusions and Overview} 
 
The Pierre Auger Observatory has been working since the start of installation in 2004, 
and has now collected an amount of data corresponding to just two years of the full Observatory but larger than the predecessor detectors. 
Using a Hybrid technique, the Observatory profits from the direct energy determination from the Fluorescence Detectors, and extends 
that precision to the full Surface Detector statistics with virtually no dependence on simulations. 
 
The fast decreasing high energy cosmic ray flux shows two features, which allow us to start attempting a coherent characterization of three regions: 
\begin{itemize} 
\item{below the {\it ankle}, the flux dominated by protons created in our galaxy and trapped by its magnetic field;} 
\item{above the {\it ankle}, the flux is dominated by extra-galactic particles. These are not necessarily protons, on the contrary, 
their interactions in air seem to have higher cross-section and/or multiplicity. 
Neutral primaries have not been observed yet, and charged ones are deviated by the galactic magnetic fields leading to a featureless sky.} 
\item{above $10^{19.5}$~eV, there is a rapid flux decrease. 
The energy is enough for the CMB to induce photo-disintegration of heavy nuclei and proton resonances, limiting the observable distance, 
but also enough for charged particles to overcame magnetic fields, allowing for charged particle astronomy in this limited part of the Universe.} 
\end{itemize} 
However, more data and more analyzes are necessary to fully fix this 
simplified picture. While the identification of individual cosmic 
ray sources will mean an important breakthrough for Astrophysics, 
Particle Physics will be necessary to interpret what exactly arrives 
to Earth. Even if the data seem to indicate we are in the presence 
of heavy nuclei primary, no model can fully explain the data today. 
A combination of variables will give more information in the near 
future. The detailed study of these particle interactions in the 
atmosphere might constrain the hadronic models evolution to energies 
far beyond the reach of man-made accelerators. 

\section{Acknowledgements}
I thank my colleagues in The Pierre Auger Collaboration for 
their carefull reading of this manuscript. 
This work is partially funded by Funda\c{c}\~ao para a Ci\^encia 
e Tecnologia, Portugal (CERN/FP/109286/2009).

\end{document}